\documentclass[12pt]{article}
\usepackage{graphics}
\let\ssection=\section
\renewcommand{\section}{\setcounter{equation}{0}\ssection}

\newcommand\mathC{\mkern1mu\raise2.2pt\hbox{$\scriptscriptstyle|$}
        {\mkern-7mu\rm C}} 
\newcommand{\mathR}{{\rm I\! R}}         

\newcommand{\W}{\ensuremath{{\cal W}}}

\renewcommand{\H}{\ensuremath{{\cal H}}}

\newcommand{\D}{\ensuremath{{\bf D}}}

\newcommand\bi{\begin{itemize}}
\newcommand\ei{\end{itemize}}

\begin{document}
\begin{titlepage}
\hspace{10truecm}Imperial/TP/99-0/11

\begin{center}
{\large\bf An obstruction based approach to the Kochen-Specker theorem}
\end{center}

\vspace{0.8 truecm}

\begin{center}
        J. ~Hamilton\footnote{email: j.hamilton@ic.ac.uk}\\[10pt]

    The Blackett Laboratory\\ Imperial College of Science,
        Technology \& Medicine\\ South Kensington\\ London SW7 2BZ\\
\end{center}

\vspace{0.8 truecm}

\begin{abstract}

In \cite{IB98} it was shown that the Kochen Specker theorem can be
written in terms of the non-existence of global elements of a certain
varying set over the category \W\ of boolean subalgebras of projection
operators on some Hilbert space \H. In this paper, we show how
obstructions to the construction of such global elements arise, and
how this provides a new way of looking at proofs of the
theorem.

\end{abstract}
\end{titlepage}

\section{Introduction}

The Kochen-Specker (KS) theorem states that there exist no valuations
on the set of self-adjoint operators on a Hilbert space \H\ of
dimension greater than two. Valuations $V$ are required to respect the
functional relationships between the operators, so if $\hat B = f(\hat
A)$, then
\begin{equation}
V(\hat B) = f (V (\hat A)).  \label{eqn:func}
\end{equation}
In particular this implies that for a projection operator $\hat P$,
$V(\hat P) = 0 $ or $ 1 $.
This requirement is usually referred to as the functional composition
principle, or $FUNC$.

Many proofs of the KS theorem are written in terms of `colouring'
rays in a Hilbert space of fixed dimension, assigning values $0$
or $1$ to collections of rays in a $n$-dimensional Hilbert space,
subject to the condition that the sum of the values of $n$
orthogonal rays is one. This is equivalent to assigning values to
the projection operators onto those rays subject to the condition
$FUNC$ above.

Since Kochen and Specker's original collection of 137 rays in three
dimensions \cite{KS67} which is not `colourable'
 in this way, many smaller sets of rays have
been discovered (see \cite{bubbook} for a review). It is perhaps
hard to see whether these sets are related, and why it is, for
example, that there exist non-colourable sets of rays in four
dimensions which are smaller than those for three dimensions. The
observation in \cite{IB98} that the Kochen-Specker theorem may be
written in terms of the non-existence of global elements of a
certain presheaf provides a new way of looking at these proofs, and
shows new relationships between them.

In this paper I will briefly review the statement of the KS theorem
in presheaf-theoretic terms before showing how this approach brings
out a different structure, based on the relationships between
operators (or their spectral algebras) rather than rays, and how this
structure may be seen in existing proofs of the theorem. The KS
theorem may be seen as the statement that there exist obstructions to
the construction of global elements of a particular varying set over
the partially ordered set \W\ of boolean subalgebras of projection
operators on a Hilbert space of dimension greater than two. The
obstructions arise over (collections of) `loops' of algebras in \W. We
show that for a Hilbert space of  three dimensions, these loops necessarily contain ten
algebras, but that this number may be reduced for higher-dimensional
Hilbert spaces.

\section{The Kochen-Specker theorem in terms of varying sets}

In \cite{IB98,IB99} it was shown that the Kochen-Specker theorem can be
stated as the fact that the dual presheaf over the category \W\ of
Boolean subalgebras of projection operators on a Hilbert space \H\
has no global elements. In this section, we describe this construction
for the case of a finite-dimensional Hilbert space.

We begin with the set \W\ of Boolean subalgebras of projection operators on
\H. This forms a partially ordered set by subalgebra inclusion.
If $W_2 \subseteq W_1$, we write the inclusion map as $i_{W_2W_1}: W_2
\rightarrow W_1$.\footnote{In this way \W\ forms a category with
objects being Boolean subalgebras and morphisms being \\
\hspace*{1.5em} inclusion maps.} 

A {\em varying set\/} ${\bf X}$ over the partially ordered set \W\ is
an assignment of a set ${\bf X}(W)$ to each $W \in \W$, and an
assignment to each inclusion map $i_{W_2W_1}: W_2
\rightarrow W_1$ of a map ${\bf X}(i_{W_2W_1}):{\bf X}(W_1)\rightarrow
{\bf X}(W_2)$  between the sets associated with $W_1$ and $W_2$. If 
 $W_3 \subset W_2 \subset W_1$, we require that ${\bf
X}(i_{W_3W_2})\circ {\bf X}(i_{W_2W_1}) = {\bf X}(i_{W_3W_1})$. 

This structure is also known as a presheaf over the {\em base
category} \W. It can be thought
of as a bundle over \W\ with extra structure, namely the maps ${\bf
X}(i_{W_2W_1})$ between `fibres' over $W_1$ and $W_2$. A {\em global
element\/} $\gamma$ of the varying set ${\bf X}$ is analogous to a
global section of a bundle; it is a function which assigns to each $W$
in \W\ an element $\gamma(W) \in {\bf X}(W)$,  
with the property that the elements match up under the action of
the varying set maps, so for $W_2 \subset W_1$, we have that
\begin{equation}
\gamma(W_2) = {\bf X}(i_{W_2W_1})(\gamma(W_1)) \label{eqn:match}
\end{equation}

The varying set ${\bf D}$ over \W\ was introduced in \cite{IB98}.
It is defined as follows:
\begin{enumerate}
\item On elements of \W: ${\bf D}(W)$ is the {\em dual\/} of $W$;
thus it is the set ${\rm Hom}(W,\{0,1\})$ of all homomorphisms
from the Boolean algebra $W$ to the Boolean algebra $\{0,1\}$.

\item On inclusion maps: if $i_{W_2W_1}:W_2\rightarrow W_1$ then
${\bf D}(i_{W_2W_1}): {\bf D}(W_1)\rightarrow {\bf D}(W_2)$ is
defined by ${\bf D}(i_{W_2W_1})(\chi):=\chi|_{W_2}$ where
$\chi|_{W_2}$ denotes the restriction of $\chi\in {\bf D} (W_1)$
to the subalgebra $W_2\subseteq W_1$.
\end{enumerate}

A global element of this varying set, if it existed, would
then be a function $\delta$ that associates to each
$W\in\cal W$ an element $\delta(W)$ of the dual of $W$ such that
if $i_{W_2W_1}: W_2\rightarrow W_1$ then
\begin{equation}
\delta(W_1)|_{W_2}=\delta(W_2).  \label{eqn:D-matching}
\end{equation}

Since each projection operator $\hat P \in {\cal P(H)}$ belongs to at
least one Boolean algebra, namely the algebra $W_P := \{\hat 0,\hat
1,\hat P,\neg\hat P \}$, and this is included in any larger algebra
$W$ which also contains $\hat P$, a global element $\delta$ assigns a
value $V_{\delta}(\hat P) := \delta(W_P)(\hat P)$ to each projector,
and this value (either $0$ or $1$) is equal to $\delta(W)(\hat P)$ for
all Boolean algebras $W$ containing $\hat P$.  Furthermore, if $\hat
P\land\hat Q=\hat 0$ there exists an algebra $W_{PQ}$
containing both $\hat P$ and $\hat Q$, and since $\delta(W_{PQ})$ is a
homomorphism, we have that
\begin{equation}
\delta(W_{PQ})(\hat P\lor\hat Q)= V_{\delta}(\hat P\lor\hat Q)
= V_{\delta}(\hat P)+V_{\delta}(\hat Q),  \label{eqn:PplusQ}
\end{equation}
and hence $\delta$ provides a valuation on projectors which respects the
functional composition principle.  The KS
theorem may therefore be stated as the fact that, for a Hilbert space
of dimension greater than two, the varying set $\bf D$  has no global elements.

 It is this type of valuation which is usually
used in the construction of counterexamples to the Kochen-Specker
theorem---sets of directions are given for which the associated
projectors cannot be assigned
the values $0$ or $1$ in accordance with the above condition. These
may be viewed as belonging to subsets of \W\ over which there is some
obstruction to constructing a global element of \D.

\section{Obstructions to the construction of a global element}

The statement of the KS theorem in this way, in terms of global
elements, provides a new insight into its origin, and how it may
be proved. We know that we can construct a  {\em partial
element\/} of \D\ over certain subsets of \W\ (cf. discussion of
partial valuations in \cite{IB98}, Section 3.2), but that these
cannot be extended over the whole of \W\ without violating the
matching condition Eq.\ (\ref{eqn:D-matching}). We now proceed to
identify possible obstructions to such a construction.

The simplest possible obstruction which could occur to the
construction of a global element $\gamma$ of a varying set $\bf X$ over a
partially ordered set $\cal P$ is if there are four {\em objects} $A,B,C,D
\in \cal P$ such that $C$ and $D$ each have a map to both $A$ and
$B$, which we denote $f_{CA}: C \rightarrow A$ , $f_{DA}:D \rightarrow
A$, etc.
%
%
\begin{equation}
\begin{array}[c]{ccccc}
        C& \rightarrow &A& \leftarrow &D\\
 &\searrow &&\swarrow
  \\ &&B&& \end{array} \label{eqn:2-loop}
\end{equation}

\begin{figure}[htb]
\begin{center}
\resizebox{0.5\textwidth}{!}{%
\includegraphics{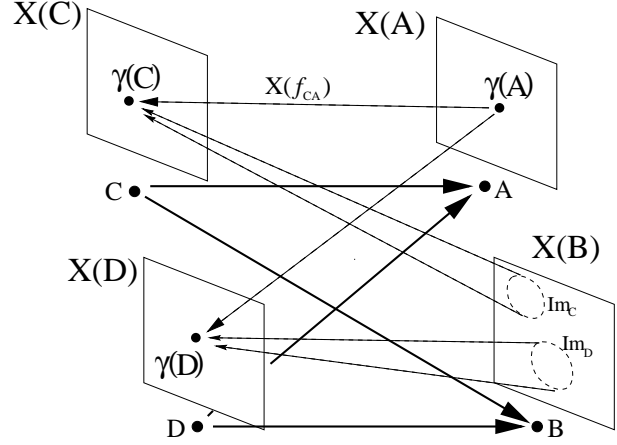}}
\caption{A possible obstruction}
\label{fig:4obs}
\end{center}
\end{figure}

If we then pick a value for $\gamma(A)$, say $\gamma(A) = x_A \in {\bf
X}(A)$, the varying set 
map ${\bf X}(f_{CA}) : {\bf X}(A) \rightarrow {\bf X}(C)$ picks a
unique element ${\bf X}(f_{CA})(\gamma(A)) =
\gamma(C) \in {\bf X}(C)$, and similarly there is a unique  $\gamma(D)
\in {\bf X}(D)$ (Figure \ref{fig:4obs}). For the matching condition
Eq.\ (\ref{eqn:match})
to hold, $ \gamma(B)$ must be in the inverse image ${\rm Im}_C =
\left({\bf X}(f_{CB})\right)^{-1}(\gamma(C)) \subset {\bf X}(B)$ of
$\gamma(C)$ along the map ${\bf X}(f_{CB})$, and also in the inverse
image ${\rm Im}_D$ of $\gamma(D)$. If ${\rm Im}_C \cap {\rm Im}_D =
\emptyset$, then this is clearly not possible
(as shown in Figure \ref{fig:4obs}). In that case, we know that there is
no global element of $\bf X$ with the chosen value $\gamma(A) = x_A$ at stage
$A$. If there is no value $\gamma(A)$ for which ${\rm Im}_C \cap {\rm
Im}_D \ne \emptyset$, then we have an obstruction, and the varying set
$\bf X$ has no global elements.

For most varying sets, such simple obstructions will not exist. More
complicated obstructions will arise over larger collections of objects
in the base category $\cal P$ in the same basic way, namely that given an
assignment of a value for the global element over one object in $\cal P$, there will be
two (or more) incompatible restrictions on the possible values the
global element may have for another object. We therefore expect
obstructions to arise over sets of objects $A_i$ where each is connected to
at least two others ({\em i.e.} there are at least two relations in
${\cal P}$  with each
$A_i$ as their domain or codomain).

The simplest structures to look for are therefore `loops' of objects
in the partially ordered set $\cal P$; each object being connected to
precisely two others. We can view Eq.\ (\ref{eqn:2-loop}) as a loop of
four objects. For the varying set $\bf D$ over \W, loops containing
just four Boolean subalgebras will not suffice to produce
obstructions. The loops needed will be larger, and many connected
loops will generally be required to produce a complete obstruction.

\section{Obstructions in a 3-dimensional Hilbert Space}

\subsection{Loops of Operators}

In the category $\W_3$ of Boolean subalgebras of  a three-dimensional Hilbert space $\H_3$, there
are three types of Boolean algebra, corresponding to operators with
different degeneracies.
\begin{enumerate}
\item {\em Maximal algebras\/}
containing three projectors $\hat P_A , \hat P_B, \hat P_C$ onto
orthogonal 1-dimensional subspaces $A,B,C$ of $\H_3$. These will be denoted
$M_{ABC} = \{ \hat P_A , \hat P_B, \hat P_C, \neg\hat P_A ,\neg \hat
P_B,\neg\hat P_C,\hat 1,\hat 0  \}$.
\item{\em Degenerate algebras\/} containing just one
1-dimensional projector $\hat P_A$. These will be denoted $L_A = \{\hat
P_A, \neg\hat P_A, \hat 1 ,\hat 0\}$.
\item{The trivial algebra\/} $\{\hat 1 ,\hat 0\}$, included in all $W$ in
\W.
\end{enumerate}

Inclusion maps in $\W_3$ (other than identity maps) have either the
trivial algebra or a degenerate algebra as their domain. Maps from the
trivial algebra are of no use in constructing obstructions;
$\delta(\hat 1) = 1$ for any global or partial element $\delta$. We
will therefore only need to consider morphisms of the type
$i_{L_AM_{ABC}}: L_A \rightarrow M_{ABC}$.

We can see straight away that there are no loops such as those in Eq.\
(\ref{eqn:2-loop}) with just two maximal algebras, $M_{ABC},M_{DEF}$
and two degenerate algebras $L_X,L_Y$ both included in each
maximal. This would require each degenerate algebra to contain a
one-dimensional projector which is in each of the maximal algebras,
with the result that either $M_{ABC}$ and
$M_{DEF}$ have two projectors in common (and hence must also share a
third, and be equal), or $\hat P_X = \hat P_Y$, and so $L_X = L_Y$.

In fact, for a three-dimensional Hilbert space, there are also no
loops with three or four maximal algebras.
In a loop with four maximal algebras, each 
must share a single projector with two other maximal
algebras. The same projector must not be shared by any three maximal
algebras, or else they will all be linked by a single degenerate
algebra, and there would  a reduced loop with fewer algebras.

So the four maximal algebras contain 8 one-dimensional  projectors,
denoted $\hat P_A,\hat P_B\ldots\hat P_G$. We use the letters
$A,B,\dots G$ to
refer to the vectors in \H\ corresponding to those projectors.

The algebras  are joined as follows:
\begin{equation}
\begin{array}{ccccccccccccccccc}
M_{ABC} & & & & M_{ADE} & &  & & M_{DFG} & & & &M_{FHB}& & & &M_{ABC} \\
&\nwarrow &  & \nearrow & &\nwarrow &  & \nearrow & &\nwarrow &  &
\nearrow &   &\nwarrow &  & \nearrow &  \\
 & &  L_A & & & & L_D & & & & L_F & & & & L_B
\end{array}
\end{equation}

Then since $\neg \hat P_A = \hat P_B \vee \hat P_C = \hat P_D \vee
\hat P_E$, we know that  $D$ is a linear combination
of $B$ and $C$, $D = x_B B + x_C C$ for some $x_B,x_C \in \mathC$.
Similarly, $\neg \hat P_F = \hat P_D \vee \hat P_G = \hat P_H \vee
\hat P_B$, and $\neg \hat P_B = \hat P_A \vee \hat P_C = \hat P_F \vee
\hat P_H$,
 so $D = y_B B + y_H H$ and $C= z_F F + z_H H$ for some $y_B,
y_H,z_F,z_H \in \mathC$. So
\begin{equation}
D =  y_B B + y_H H = x_B B + x_C C = x_B B + x_C (z_F F + z_H H)
\end{equation}
but we know that $B,H$ and $F$ are mutually orthogonal, so
\begin{equation}
 y_B B + y_H H =  x_B B + x_C (z_F F + z_H H)
\end{equation}
implies that $ x_C z_F = 0$, and hence either $D=B$ or $C=H$, and it
quickly follows that all of the maximal algebras are equal.

Therefore in a three-dimensional Hilbert space, there are no loops in $\W_3$
containing just four maximal algebras.

\subsection{Definite Prediction Sets and Loops in 3 Dimensions}

Although loops in $\W_3$ with five maximal algebras do exist, a single
such loop (with ten associated vectors) does not provide a complete
obstruction; partial elements over such loops certainly exist. To form
a complete obstruction, several such loops must be joined together.
To see how obstructions may be built up in this way, we will now look
at the structures corresponding to some existing proofs of the KS
theorem.

Currently, most proofs in three dimensions consist of the construction
of a set of rays in $\H_3$ with the property that they cannot each be
assigned a value $0$ or $1$ subject to the constraint that the sum of
values assigned to any orthogonal triple of rays is one. This is
usually described in terms of `colouring' rays, green for zero and red
for one. Following terminology in \cite{CG95} such a set of vectors is
called a {\em totally non-colourable set\/}, or TNCS.

One way to produce such a TNCS is that used originally by Kochen and
Specker, whereby a TNCS are
built up from sets with less predictive power.
The starting point is a
{\em definite prediction set\/} (DPS), which is a set of rays $\{ r_i
\}$ such that if a single\footnote{Cabello and Garc\'{\i}a-Alcaine
\cite{CG95} also use the term DPS for a construction whereby more than one input
is required for a prediction.}  particular `input value' is chosen for
one ray, $r_j$, there is another vector, $r_k$, which is not orthogonal to
$r_j$, for which there can be only one possible value.

New DPSs may be created by rotating each ray in the original set by
the same
fixed amount, preserving their orthogonality relations.
One can then proceed to
chain several such DPSs together, with the predicted value of the
first being used to constrain the `input
value' of the second, and so on, until a structure is obtained
where the final predicted value contradicts the initial input
value. This results in a partially non-colourable set (PNCS); there is no
possible colouration consistent with the initial input value. Such
sets constitute state-dependent proofs of the KS theorem---if the
system is in an eigenstate corresponding to the initial input value, a
contradiction is seen to occur.

A TNCS can then be produced from PNCSs; in three dimensions this
would require three PNCSs, one to eliminate each possible
valuation on an orthogonal triple of rays.

Clifton \cite{Cl92} showed how a DPS may also be used to directly give a
proof of the need for values to be contextual. This involved also
using some arguments based on the statistical predictions of quantum
mechanics.

\subsubsection{Clifton's DPS}

The DPSs used by Clifton, Cabello and Garc\'{\i}a-Alcaine and Kochen
and Specker may be
easily shown to consist of vectors corresponding to the one
dimensional projectors in certain algebras which
form small collections of interlocking loops in $\W_3$. Clifton uses a set of
eight vectors
\begin{displaymath}
\begin{array}{rlcrlcrl}
r_1 =&  (1,0,-1) & &
r_4 =&  (0,0,1) & &
r_7 =&  (1,1,1)     \\
r_2 =&  (1,0,1) & &
r_5 =&  (1,-1,0) & &
r_8 =&  (-1,1,1)     \\
r_3 =&  (0,1,0) & &
r_6 =&  (1,1,0) & & &  \\
\end{array}
\end{displaymath}
consisting of two orthogonal triples, $\{r_1,r_2,r_3\}$ and  $\{r_4,r_5,r_6\}$
plus two other vectors, each of which is
orthogonal to one vector in each orthogonal triple.

\begin{figure}[htb]
\begin{center}
\scalebox{.8}{%
\includegraphics*[4cm,18cm][13cm,26cm]{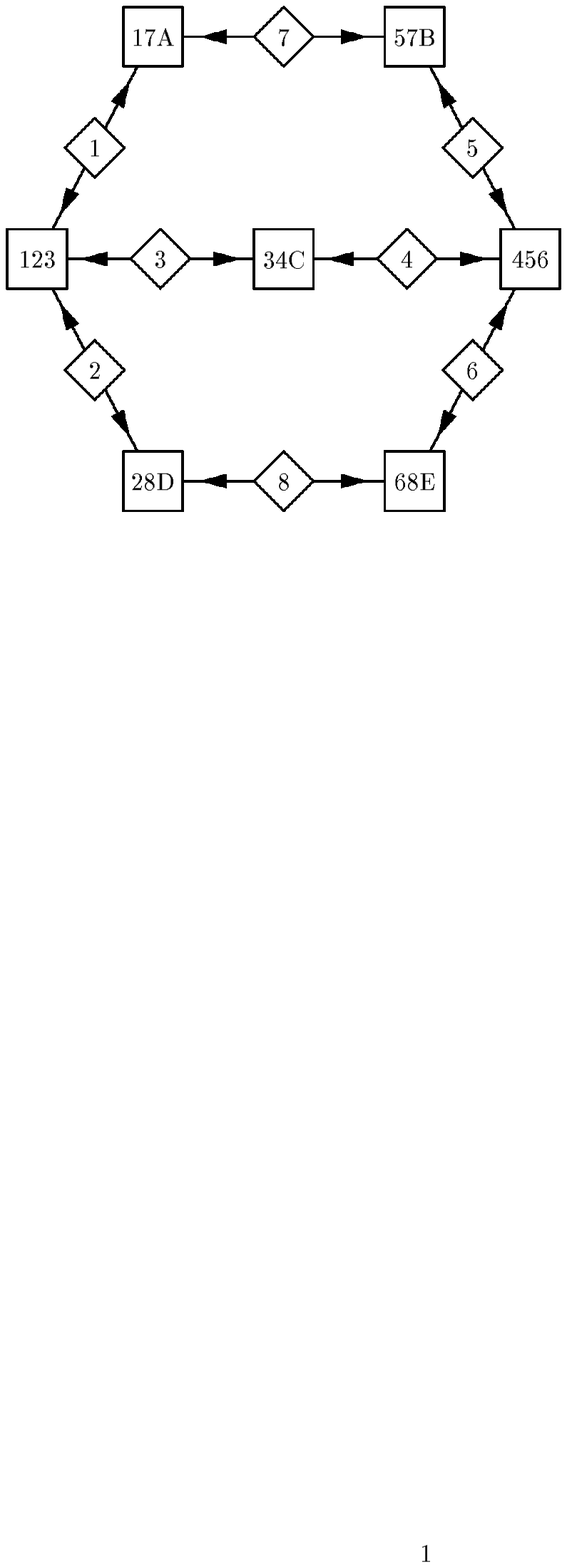}}
\caption{Clifton's DPS as a subset of \W}
\label{fig:7dps}
\end{center}
\end{figure}

These vectors fit into seven algebras, as shown in Figure
(\ref{fig:7dps}). The eight diamonds represent degenerate algebras,
those generated by projectors onto the above rays. They are labelled
by the number of the appropriate vector from the above set. The
squares represent maximal algebras, each containing three projectors
onto one-dimensional rays, plus all meets and joins. These are
labelled by the numbers of the appropriate vectors from the above set,
The letters $A,B\ldots E$ correspond to new rays not in the set
used by Clifton. However, since each new ray is orthogonal to two in
Clifton's set, a valuation on Clifton's set will also completely
determine the value of these new rays.

Figure (\ref{fig:7dps}) is seen to form a DPS in the following
way. Suppose we assign value one to the projectors onto the rays $r_7$, and $r_8$.
\begin{equation}
V(\hat P_7) =  V(\hat P_8) = 1     \label{eqn:dpsinit}
\end{equation}
Then $V(\neg \hat P_7) = 0$. This implies that $V(\hat P_1) =
V(\hat P_5) = 0$ through the application of Eq.\ (\ref{eqn:PplusQ}) in
the top two maximal algebras in the diagram. Similarly,  $V(\neg \hat P_8) = 0$
implies that $V(\hat P_2) = V(\hat P_6) = 0$.

In algebra $M_{123}$ we now have $V(\hat P_1) = V(\hat P_2) =
0$, which implies that $V(\hat P_3) = 1$, and similarly in algebra
$M_{456}$ we can deduce that $V(\hat P_4) = 1$. This gives us a
contradiction in algebra $B$ --- two orthogonal projectors $\hat
P_{3}$ and $\hat P_{4}$ are both assigned the value $1$. Hence by
{\em reductio ad absurdum\/} the value assignment in Eq.\
\ref{eqn:dpsinit} is not possible, and we deduce that if $V(\hat
P_{7}) = 1$, then we have the `definite prediction' that we must assign $V(\hat P_8) =0$.

\subsubsection{Cabello and Garc\'{\i}a-Alcaine's DPSs}

The construction above, Figure (\ref{fig:7dps}), with seven maximal
algebras forming two loops of five,  with three algebras in both loops, is
the smallest way  to join two loops together. As has been shown, this
leads to a DPS where, if a certain 1-dimensional projector is assigned
value 1, another such must be assigned the value zero. In \cite{CG95},
Cabello and Garc\'{\i}a-Alcaine give three examples of DPSs with a
stronger predictive power, namely that if a certain 1-dimensional
projector is assigned value one, then another 1-dimensional projector
must also be assigned value one (rather than zero, as in the Clifton
DPS). The advantage of their construction is that chains of such DPSs
may be formed: a new DPS is constructed  by rotating all directions by
a fixed amount so that the direction of the ray whose value is predicted
by the first set becomes the direction for input value for the second.

For the three-dimensional case, they give a set of ten vectors as
follows:
\begin{displaymath}
\begin{array}{clccl}
r_1 =&  (1,0,0)    &  &
 r_6 =&  ({\rm cot} \phi,1,-{\rm cot}\beta)      \\
r_2 =&  (0,{\rm cos} \alpha,{\rm sin}\alpha)    &  &
r_7 =& ({\rm tan}\phi {\rm cosec}\beta,-{\rm sin}\beta,{\rm cos}\beta)   \\
r_3 =&  ({\rm cot} \phi,1,-{\rm cot}\alpha)    &  &
 r_8 =&  ({\rm sin}\phi,-{\rm cos}\phi,)       \\
r_4 =&  ({\rm tan}\phi {\rm cosec}\alpha,-{\rm sin}\alpha,{\rm cos}\alpha)&&
r_9 =&  (0,0,1)       \\
r_5 =&  (0,{\rm cos} \beta,{\rm sin}\beta)   &  &
r_{10} =&  ({\rm cos}\phi,{\rm sin}\phi,0)
\end{array}
\end{displaymath}
with $\mid\phi\mid \le {\rm arctan}(1/\sqrt{8})$, $\alpha \ne \beta$,
and $\alpha$, $\beta$ and $\phi$ related by:
\begin{equation}
{\rm sin}\, \alpha\;{\rm sin}\, \beta\;{\rm cos}\,(\alpha - \beta) = - {\rm tan}^2\phi.
\end{equation}

\begin{figure}[htb]
\begin{center}
\scalebox{0.7}{%
\includegraphics*[4cm,14.5cm][17cm,26cm]{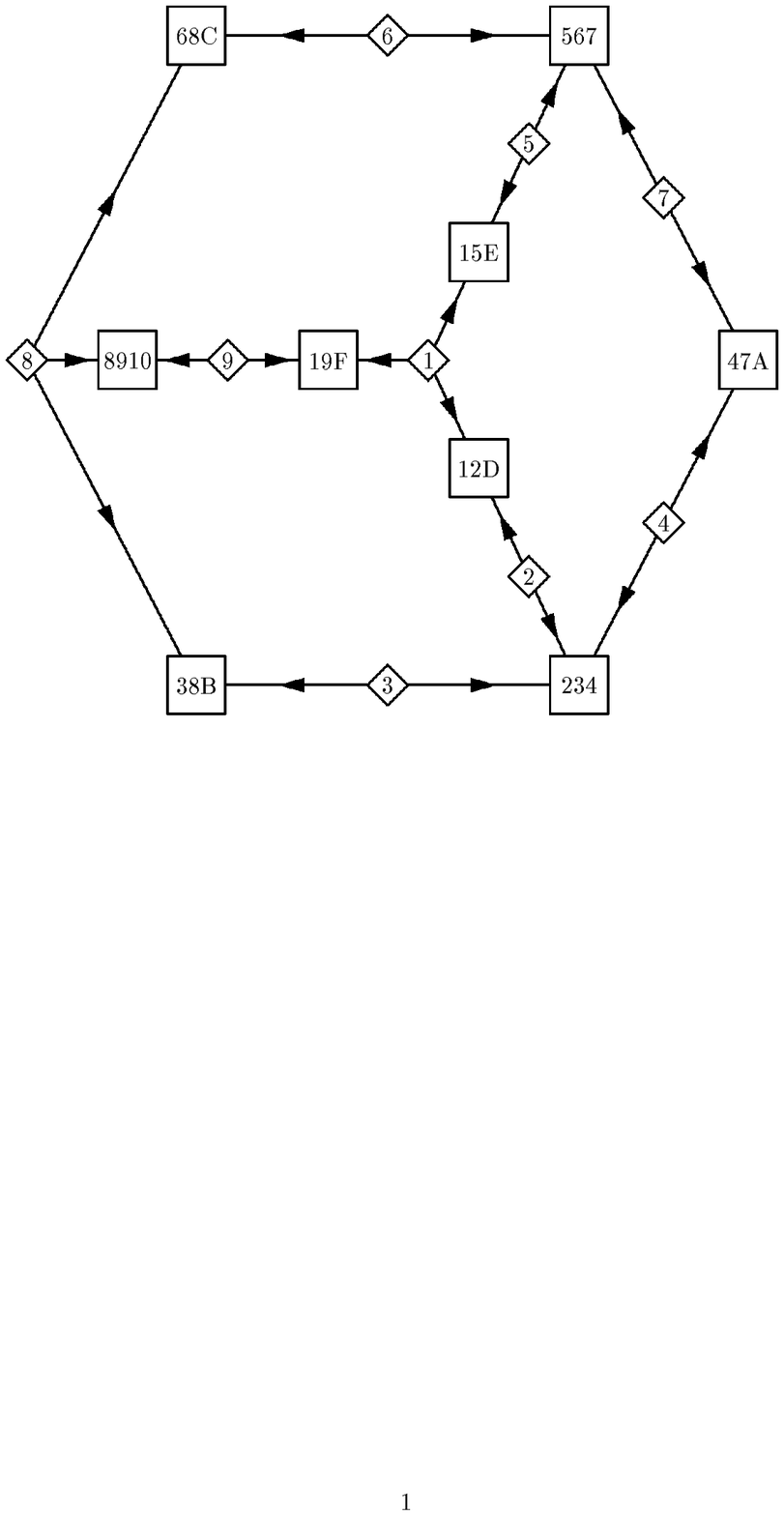}}
\caption{A DPS in $\W_3$ used by Cabello and Garc\'{\i}a-Alcaine}
\label{fig:9dps}
\end{center}
\end{figure}

Cabello and Garc\'{\i}a-Alcaine give a diagram showing the
orthogonality relations of these vectors. They can also be represented
as a diagram in \W, as shown in Figure (\ref{fig:9dps}). The degenerate
algebras are labelled by the number (in the above list) of the single
direction onto which they contain a one-dimensional projector. The maximal
algebras are labelled by the three directions they contain. As in the case
of Clifton's DPS, some new directions $A,B\ldots F$ are required, each
orthogonal to two from the above set.

 This construction can be seen to contain the same structure as
the DPS due to Clifton (two loops of five maximal algebras with
three in common---the right hand and outer loops), with an extra
connection formed by the maximals containing the directions
$\{8,9,10\}$ and $\{1,9,F\}$. The prediction used by  Cabello and
Garc\'{\i}a-Alcaine in this set is that if the projector onto $r_1$
is assigned the value one, then the value assigned to $r_{10}$ must
also be one. This can again be seen by assuming $V(\hat P_{10}) =
0$ and following the implications round to reach a contradiction.

\begin{figure}[htb]
\begin{center}
\scalebox{0.5}{%
\includegraphics*[3cm,14cm][28cm,26cm]{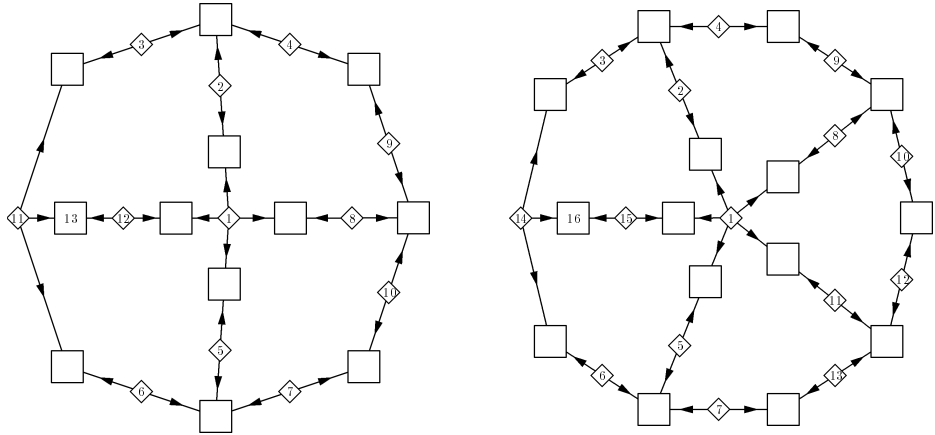}}
\caption{Two more DPSs in $W_3$}
\label{fig:2a3}
\end{center}
\end{figure}

Cabello and Garc\'{\i}a-Alcaine also identify two more DPSs
(\cite{CG95}, Appendix) which correspond to the diagrams in \W\ shown
in Figure (\ref{fig:2a3}).

\subsection{Totally Non-Colourable Sets}

Complete geometrical proofs of the Kochen-Specker theorem in three
dimensions consist of large numbers of algebras, connected in an
intricate way. As has been mentioned, this may be done by chaining
together DPSs such as those shown in the previous Subsection.
However, the number of rays can be reduced by considering sets of
rays with more interconnections.

\begin{figure}[htb]
\begin{center}
\scalebox{.8}{%
\includegraphics*[4.5cm,10.5cm][19cm,26cm]{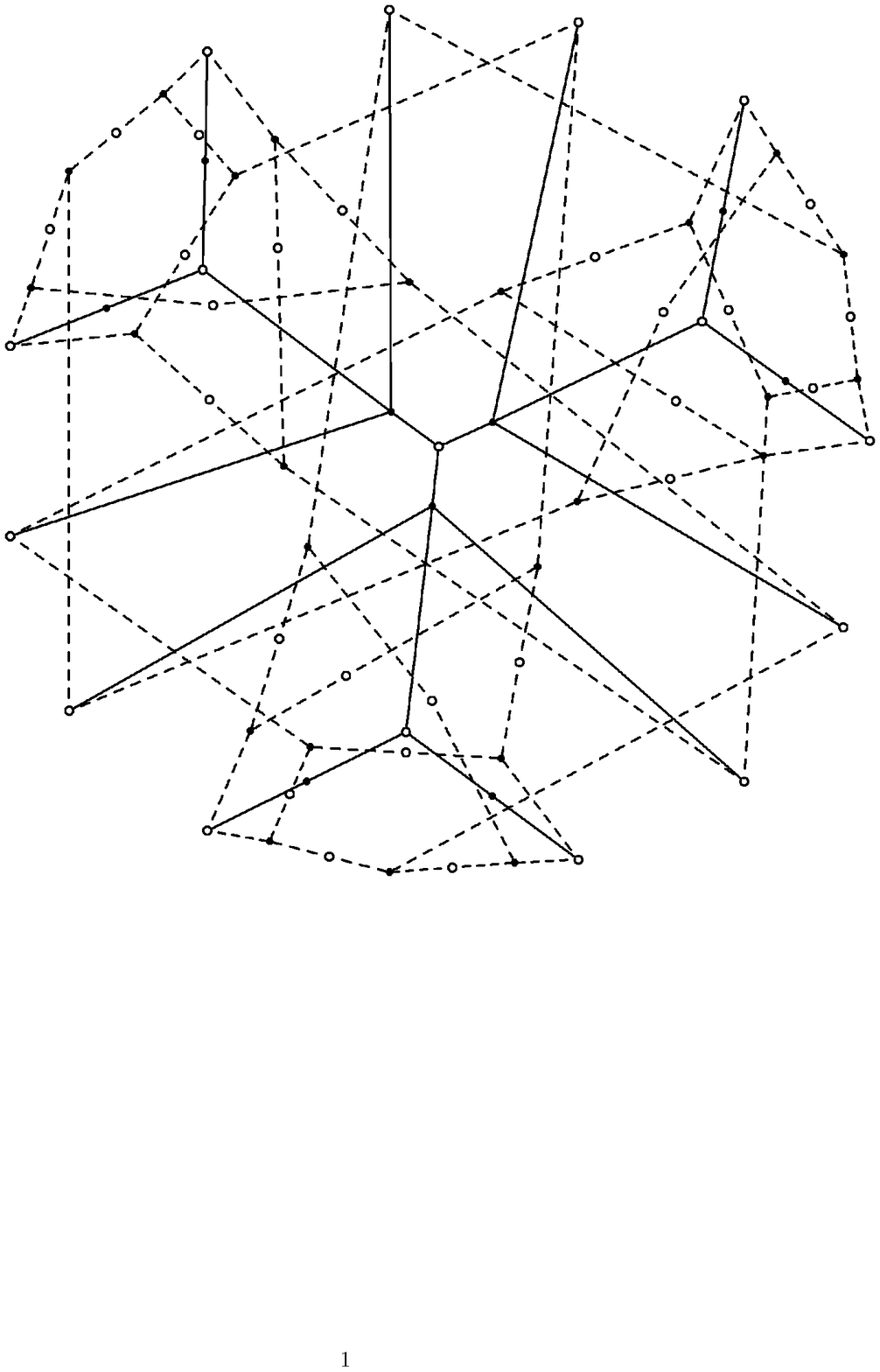}}
\caption{Peres' 3-d TNCS}{Maximal algebras are shown as circles,
degenerate algebras as solid black dots.}
\label{fig:big}
\end{center}
\end{figure}

For example, the proof by Peres \cite{Pe90}, a TNCS with 33 rays, can
be drawn as a diagram in \W\ with 40 maximal algebras. To do this, 24
more rays must be introduced, each of which is orthogonal to two
others in the set of 33, and each of which is only contained in one
maximal algebra. The diagram produced in this way (Figure
\ref{fig:big}) is large, having many interlinked loops of algebras,
although as expected, loops containing just five maximal algebras may
be identified within it. The threefold symmetry within the diagram
comes from the fact that the set of rays used is invariant under
interchange of any two axes---the central maximal algebra contains
projectors onto the rays $(0,0,1),(0,1,0)$ and $(1,0,0)$.

\section{Obstructions in a Hilbert space $\H_4$ of four dimensions}

In a four-dimensional Hilbert space, there exist more types of Boolean
sub-algebras of ${\cal P(H}_4)$, corresponding to operators with different degeneracies.
In addition to the trivial algebra $\{\hat 1 , \hat 0 \}$, we have
\begin{enumerate}
\item {\em Maximal algebras\/}
containing four projectors  onto
orthogonal 1-dimensional subspaces of $\H_4$.
\item{\em 2-1-1 degenerate algebras\/} containing two 1-dimensional
projectors  $\hat P_A , \hat P_B$. These correspond to operators with
three eigenvalues, one of which is degenerate (associated with the
two-dimensional eigenspace orthogonal to the $A$ and $B$ directions).
\item{\em 3-1 degenerate algebras\/} generated by a single
one-dimensional projector.
\item{\em 2-2 degenerate algebras\/} containing  two (orthogonal)
two-dimensional projectors, and no one-dimensional ones.
\end{enumerate}

This larger number of possible types of algebra results in more
possible loops of algebras; in particular, it is possible to form
loops in
four dimensions with less than the ten algebras needed in the
three-dimensional case. The constructions required to show a
contradiction of the KS theorem may be simpler, containing far fewer
algebras than necessary for the three-dimensional case.

\subsection{Extension of Proofs in Three Dimensions.}

A DPS in a three-dimensional Hilbert space such as those shown in Figs.\
(\ref{fig:9dps}) and (\ref{fig:2a3}), where an initial assignment of
value one to a one-dimensional projector induces another such
assignment to a non-orthogonal projector, can be easily generalised to
four (or more) dimensions.

The vectors involved in the construction of the old DPS, say that in  Figure
(\ref{fig:9dps}), are thought of as
belonging to a three-dimensional subspace of $\H_4$, which may be
chosen as that
orthogonal to the vector $(0,0,0,1)$. Each algebra in
Figure (\ref{fig:9dps}) is then extended by the addition of the projector
 $\hat Q$ onto the direction $(0,0,0,1)$, plus the meets and joins of
 $\hat Q$ with the other projectors in the algebra. The result is that
we may view the diagram in Figure (\ref{fig:9dps}) as a diagram of
sub-algebras of a four-dimensional Hilbert space, with squares once
again corresponding to maximal algebras, and diamonds corresponding
to 2-1-1 degenerate algebras. There should then also be an additional
3-1 degenerate algebra, $\{ \hat Q, \neg\hat Q ,\hat 1 ,\hat 0)$, which,
like the trivial algebra, is included in every other algebra in the
diagram.

The reasoning showing that the diagram is a DPS still holds true; we
now have a DPS in four dimensions. It is this extendibility of a DPS
which Cabello and Garc\'{\i}a-Alcaine use to provide a construction of
a totally non-colourable set of rays in any Hilbert space of
finite dimension of three or greater.

\subsection{Peres and Mermin's Proof}

Perhaps the simplest and most elegant proof of the KS theorem is
the four-dimensional proof produced by Peres \cite{Pe90} and
Mermin \cite{Me90}. Mermin gives a collection of nine operators in
$\H_4$ of the form $\hat A \otimes \hat B$, where $\hat A$ and
$\hat B$ are either Pauli matrices or the (two-dimensional)
identity. These are reproduced in the body of Table
(\ref{tab:4d}). Each row and column contains three operators
having eigenvalues $\pm 1$ which commute. The products of the
three operators in each row and column is the identity operator, with the
exception of the third column, whose product is minus the identity. A
valuation would consist of an assignment of a value ($0$ or $1$)
to each observable such that these functional relationships hold,
{\em i.e.}, the product of the values in each row and column would
be $+1$, except the final column which would give $-1$. This is
easily seen to be impossible, hence the KS theorem in four
dimensions is proved.

\begin{table} \label{tab:4d}
\begin{displaymath}
\begin{array}{c||c|c|c}
 &
\begin{array}{l} {\bf A_1} = (1010) \\ {\bf A_2} = (\bar{1}010) \\ {\bf A_3} = (0101) \\  {\bf A_4} = (010\bar{1})
\end{array} &
\begin{array}{l} {\bf B_1} = (1100) \\ {\bf B_2} = (1\bar{1}00) \\ {\bf B_3} = (0011) \\  {\bf B_4} = (001\bar{1})
\end{array} &
\begin{array}{l} {\bf C_1} = (1001) \\ {\bf C_2} = (100\bar{1}) \\ {\bf C_3} = (0110) \\  {\bf C_4} = (01\bar{1}0)
\end{array}
\\ \hline\hline
\begin{array}{l} {\bf D_1} = (1000) \\ {\bf D_2} = (0100) \\ {\bf D_3} = (0010) \\  {\bf D_4} = (0001)
\end{array} &
1 \otimes \sigma_z
& \sigma_z \otimes 1 & \sigma_z \otimes \sigma_z  \\ \hline
\begin{array}{l} {\bf E_1} = (1111) \\ {\bf E_2} = (\bar{1}1\bar{1}1) \\
{\bf E_3} = (\bar{1}11\bar{1}) \\ {\bf E_4} = (11\bar{1}\bar{1})
\end{array} &
\sigma_x \otimes 1 & 1 \otimes \sigma_x & \sigma_x \otimes \sigma_x
\\ \hline
\begin{array}{l} {\bf F_1} = (\bar{1}111) \\ {\bf F_2} = (1\bar{1}11) \\
{\bf F_3} = (11\bar{1}1) \\ {\bf F_4} = (111\bar{1})
\end{array} &
 \sigma_x \otimes \sigma_z & \sigma_z \otimes  \sigma_x  &
\sigma_y \otimes \sigma_y
\end{array}
\end{displaymath}
 \caption{Operators and Vectors for a
four-dimensional Proof of KS}
\end{table}

\begin{figure}[htb]
\begin{center}
\scalebox{0.5}{%
\includegraphics*[4cm,10cm][20cm,26cm]{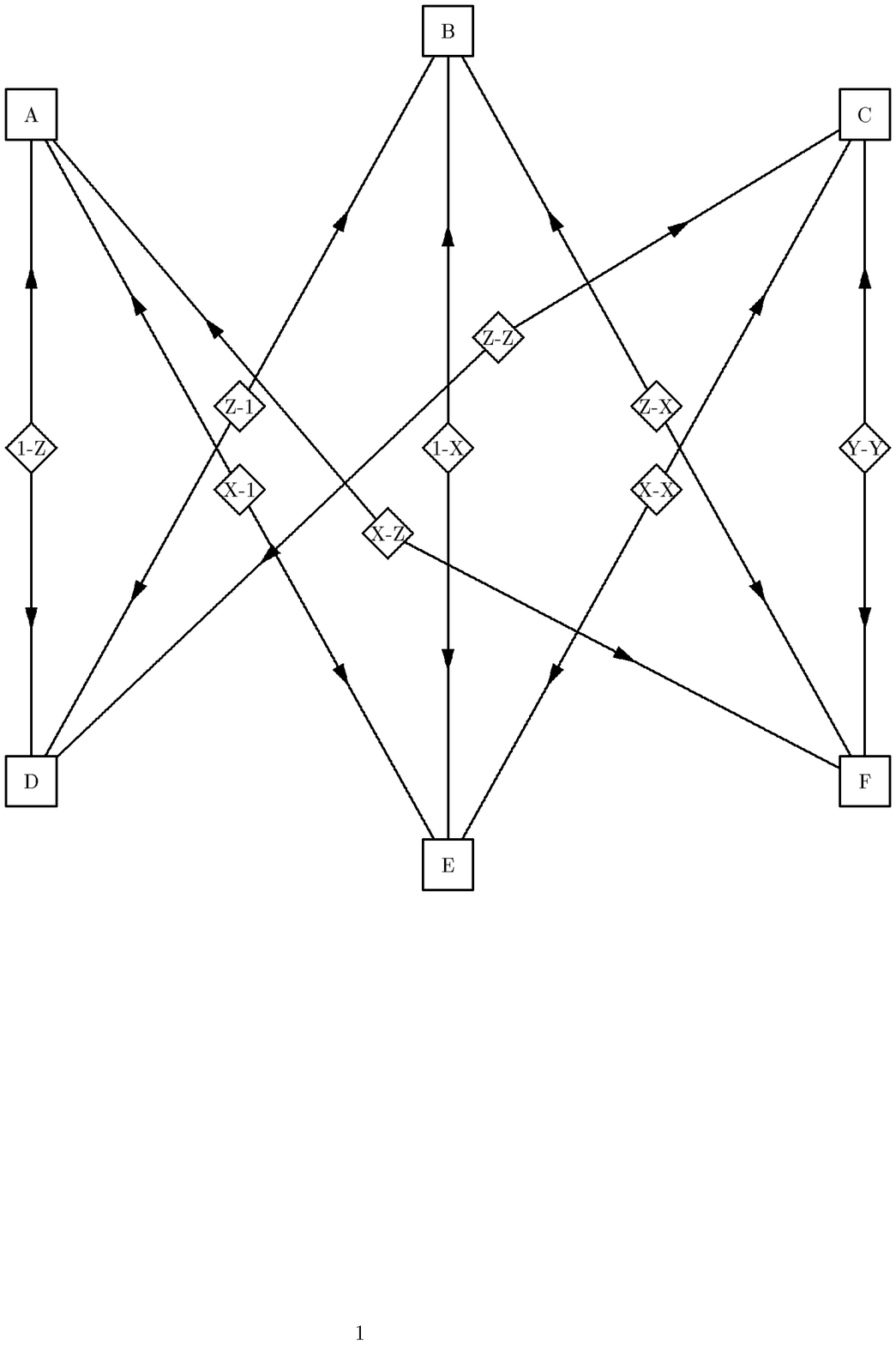}}
\caption{A TNCS in 4 dimensions}
\label{fig:4d}
\end{center}
\end{figure}

Peres gives a set of 24 vectors belonging to 6 orthogonal tetrads in
$\H_4$ (shown in the upper row and leftmost column of Table 1
which form a totally non-colourable set in the usual
way, and shows how the degenerate operators used in Mermin's reasoning may
be derived from them. When viewed in the context of constructing a
global element of \D\ over \W, it is easy to see that these correspond
to the subdiagram of \W\ shown in Figure (\ref{fig:4d}).  The orthogonal
tetrads of vectors correspond to maximal algebras $A,B\ldots F$, and the
degenerate operators in the body of the table correspond to 2-2
degenerate algebras. Each degenerate algebra is included in two
maximal algebras, for example the spectral algebra of $1 \otimes
\sigma_z$ consists of projectors onto the eigenspaces $(A_1 \otimes
A_2) = (D_1 \otimes D_3)$ and $(A_1 \otimes A_2) = (D_1 \otimes D_3)$,
where $A_i$ and $D_i$ are the vectors defined in the table.

We can see that in this four-dimensional case, there are loops containing just four maximal
algebras, and four 2-2 degenerate algebras. These constrain the
algebras involved more severely than the larger loops previously
considered, as is shown by the fact that only three interlinked
loops are now required for a complete proof of the KS theorem.

\section{Conclusions}

We have shown that various existing proofs of the Kochen-Specker
theorem may be viewed as subsets of the category \W\ over which no
global elements of the dual presheaf \D\ may be constructed. 
As we move to higher-dimensional Hilbert spaces, we see that
there are more possible types of degenerate Boolean subalgebra,
and hence more ways of forming obstructions.

We have seen that in four dimensions, loops may be easily found
containing fewer than the necessary ten algebras in three
dimensions, leading to new predictive sets. Some care must be
taken when examining these structures, however, as it is now
necessary to examine whether connections are independent. For
example, a loop with just four algebras may be constructed: two
maximal algebras, $M_{ABCD}, M_{ABEF}$ with two projectors $\hat
P_A, \hat P_B$ in common, and the two 3-1 degenerate algebras
generated by $\hat P_A $ and $ \hat P_B$. These appear to form a
loop which could be used in the construction of an obstruction,
but in fact this is not so, since the inclusion maps from both
degenerate algebras factor through the 2-1-1 degenerate algebra
generated by the projectors $\hat P_A $ and $ \hat P_B$. Any valuation
on this 2-1-1 degenerate algebra will induce (compatible) valuations on
all of its subalgebras. A choice of a value for a global element
$\gamma$ of
$\bf D$ for one of the maximal
operators will therefore only result in a single restriction on
$\gamma$ over the second, so it  is
therefore not possible to obtain contradictions from this
construction.

This fact, that loops with two degenerate algebras each included in
two maximal algebras,
may be reduced to the inclusion of a single degenerate algebra, is true
in any finite-dimensional Hilbert space. All projectors in each
degenerate algebra are contained in both maximal algebras, so a larger
degenerate algebra may be formed by combining the original two, and
this will also be contained in each maximal algebra. 

The general appearance of these diagrams of loops in \W\ suggests that
some type of cohomological description of these structures may be
possible, in analogy with the way obstructions to the construction of
non-trivial principle fibre bundles are classified.  The non-existence
of valuations would then be verifiable from properties of the base
space \W.

This would be particularly interesting in the light of recent work
by Meyer, Kent and Clifton \cite{Me99,Ke99,CK99} concerning the
fact that the KS theorem does not hold if we restrict attention to
a countable dense subset of observables. This amounts to changing
the base category \W\ from being the collection of all Boolean
subalgebras of $\cal P(H)$ to some subset of these, with the
result that over the new base category, global elements {\em can\/} be
constructed. For the construction given by Kent and Clifton in
\cite{CK99}, the reasons for the lack of obstructions are fairly
clear. They construct a subset  of projectors ${\cal P}_d
\subset \cal P(H)$ with the property that no projector in  ${\cal
P}_d$ belongs to two {\em incompatible\/} resolutions of the
identity, where two resolutions of the identity $\sum_i \hat P_i =
\sum_j \hat P_j' = \hat 1 $ are incompatible unless $[\hat
P_i,\hat P_j'] = 0 $ for all $i,j$. The corresponding category of
Boolean subalgebras, ${\cal W}_d$ will therefore contain no
degenerate algebras which are included in more than one maximal
algebra, and hence will contain no loops of algebras of the type
used in creating obstructions to the construction of global
elements. 

The subset of projectors used by Meyer's construction
\cite{Me99} ---essentially directions in $\mathR^3 \cap Q^3$
---does not have this property, and the lack of obstructions is
harder to explain. A full cohomological decription of obstructions to
the construction of global elements of presheaves would be able to
throw more light on this problem.

\end{document}